\documentclass[twocolumn,aps,prl,showpacs,preprintnumbers,superscriptaddress]{revtex4}
\usepackage{amssymb}
\usepackage{graphicx}

\begin{document}

\title{A revisitation of the 1888 H.Hertz experiment}

\author{Daniele Faccio}
\email{daniele.faccio@uninsubria.it} \affiliation{INFM and Department of Physics \& Mathematics, University
of Insubria, Via Valleggio 11, 22100 Como, Italy}
\author{Matteo Clerici}
\affiliation{INFM and Department of Physics \& Mathematics, University
of Insubria, Via Valleggio 11, 22100 Como, Italy}
\author{Davide Tambuchi}
\affiliation{ITIS Magistri Cumacini,  Via Cristoforo Colombo 1, 22100 Como, Italy}

\begin{abstract}
We propose a revisitation of the original experiment performed by H. Hertz in 1888. With a simple setup it is possible to produce electromagnetic waves with a frequency in the range of 3 MHz. By performing Fourier analysis of the signal captured by a resonant antenna it is possible to study the behaviour of the RLC series circuit, frequency splitting of coupled resonances and finally the characteristics of the near-field emitted by the loop antenna.
\end{abstract}


\maketitle
Heinrich Hertz is best known for his series of experiments conducted from 1886 onwards with which he demonstrated that the predictions of C. Maxwell were correct \cite{buchwald,adawi}. Indeed he succeeded in showing that it is possible to generate electromagnetic (EM) waves, that these propagate in free space with a well defined oscillation frequency and wavelength, that is possible to observe interference between these waves and most importantly, that these transport energy. Within the same group of experiments H. Hertz also experimentally observed for the first time many other effects, most notably the photoelectric effect.  The incredible importance and impact of these experiments is therefore clear and needs no further comment. Yet, notwithstanding this importance it is fairly rare to find these experiments reproduced in some form or another.  The scope of this paper is to give a description of the possibilities offered by an experimental setup that represents a variation of the original ``Hertz'' setup but is able to give an insight and a direct measurement of many aspects related to EM emission. Among these a clear demonstration of power transmission, a precise characterization of a loop antenna emission, the observation of frequency splitting due to resonator coupling and finally the near-field decay of the electric and magnetic fields created by a loop antenna.\\
\begin{figure}
\includegraphics[angle=0,width=7.5cm]{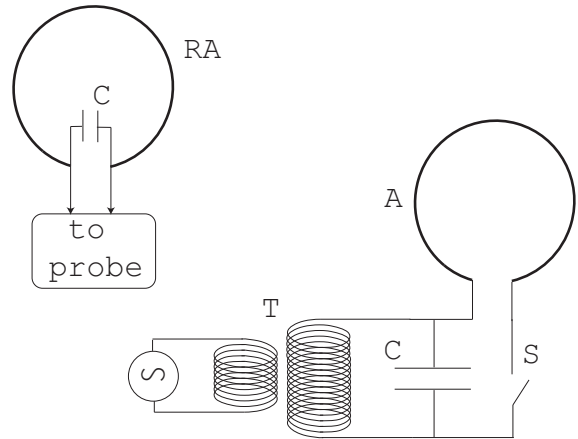}
\caption{\label{fig:fig1} Experimental layout of the experiment: A = emitting antenna (1 m diameter copper-tube loop), C = 1 nF capacitor, S = spark switch, T = 220V/50Hz to 6kV/50Hz transformer, RA = receiving antenna}
\end{figure}
The setup we used is a very simple RLC circuit designed as shown in Figs.\ref{fig:fig1} and \ref{fig:fig1bis}. A capacitance C of 1 nF (15 kV maximum voltage, from RS components) is connected to 1 m diameter loop antenna (A) obtained by bending a simple piece of copper tubing (1 inch diameter). The inductance of the antenna L is one of the quantities that we will measure with the setup: we found (see below) roughly 5 $\mu$F yet we noticed that using different diameter tubing does not change greatly this value. The resistance of this circuit is provided directly by the circuit itself (e.g. the antenna has R = 1.2 $\Omega$). The whole circuit is powered by a 6kV transformer (T). This may be found from a neon-light dealer  at a relatively low cost. Finally an important part of the circuit is the spark-switch (S) inserted in one of the arms between the capacitor C and the antenna A. This was constructed by taking two rounded bolts with a housing that allowed to regulate the distance between the rounded extremities (see the inset to Fig.\ref{fig:fig1bis}). The 6 kV voltage supply oscillates at 50 Hz. As the voltage on the capacitor increases, it does so also between the two extremities of the bolts. The breakdown threshold in air is $\sim3$ kV/mm so that if the air gap is correctly adjusted a spark will close the RLC circuit once a voltage difference of $\sim6$ kV is reached. The circuit will then start to oscillate at a frequency given by $1/2\pi\sqrt{LC}\sim2$ MHz \cite{halliday}. At each oscillation a large percentage of the power (of the order of 30\%) will be lost due to emission from the loop antenna, i.e. in the generation of propagating EM waves. These waves may then be recaptured using a second loop antenna (RA) identical to the emitting antenna A and with a series capacitor of the same value as C. As a closing comment on the setup we note that when using the 15 kV supply problems may arise with the capacitor. Indeed the maximum voltage rating for these is typically 15 kV or lower so that it is necessary to use at least two in series (and another two in parallel so as to maintain the same effective capacitance). We also had to insert the capacitors in a plastic bottle (actually a large-size soft-drink cup) filled with oil in order to avoid dielectric breakdown in the air gap between the wires protruding from the capacitors in order to prevent the latter from being burnt by the spark.\\
\begin{figure}
\includegraphics[angle=0,width=7.5cm]{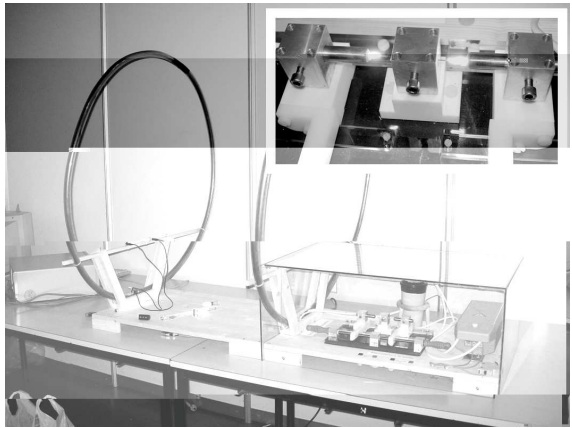}
\caption{\label{fig:fig1bis} Photograph of the experimental setup described in Fig.\ref{fig:fig1}. The inset shows a detail of the spark switch (S).}
\end{figure}
\begin{figure}
\includegraphics[angle=0,width=7.5cm]{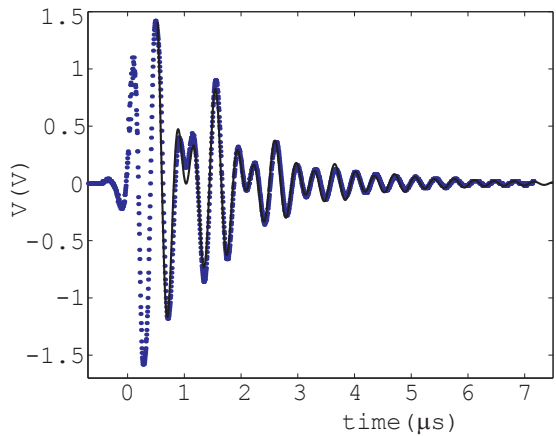}
\caption{\label{fig:fig2} Measured electric potential V versus time (dots) and best fit obtained using eq.(4) (solid line).}
\end{figure}
As a first experiment it is possible to prove energy transport in a very simple way by attaching a small light bulb to the receiving antenna RA. This experiment is best performed using the 15 kV transformer. By placing RA at distance of the order of 1 m or slightly more it is possible to light the bulb with a certain efficiency. Although the light tends to flicker we noted a great improvement and a much better stability by using a spark switch divided into two, i.e. we used three bolts in series so that a spark was generated over the two separating air-gaps. \\
For a quantitative analysis of the EM emission we attached RA to a digital oscilloscope (e.g. Tektronics TDS 1002) with which we were then able to download the acquired data to a PC (note: do not keep any laptops or  LCD  screens between or near the antennas).
For the RLC  circuit the charge on the capacitor is given by \cite{halliday}
\begin{equation}
q=Q\cdot e^{-\frac{R}{2L}t}\cos(2\pi\nu^{\prime}t+\phi)
\end{equation}
with $\nu^{\prime}=\sqrt{1/LC-(R/2L)^{2}}$. In Fig.\ref{fig:fig2} the dots show an example of an experimental V(t) trace obtained by measuring the voltage difference on the capacitor with RA placed at 1 m from A. A short build-up time is followed by the expected exponential decrease yet the oscillation does not follow a perfect cosine function. Indeed, by taking the Fourier transform of the V(t) trace we observe two distinct frequency peaks at $\nu_{1}=1.95$ MHz and $\nu_{2}=2.88$ MHz, as shown in Fig.\ref{fig:fig3}. So the shape of the V(t) trace is due to a beating between these two different frequencies. In Fig.\ref{fig:fig2} we may also see that this beating dies out after $t\sim4$ $\mu$s leaving a sinusoidal oscillation at $\nu^{\prime}=\nu_{2}$.  The physical reason for this behaviour lies in the fact that during the first oscillations of the circuit a very large EM field is emitted. This is then captured by the receiving antenna RA which may in turn re-emit thus modifying the behaviour of the RLC circuit. In other words the inductances of the two (receiving and transmitting) loop antennas  will couple. As the oscillations die away the emitted field becomes much smaller, returning the circuit to the ideal RLC state. We note that this behaviour is analogous to that observed with Tesla coils. In this case a very similar beating between two slightly different oscillation frequencies is observed and is ascribed to the periodical coupling back and forth of energy between the primary and secondary coils. Indeed the apparatus described here is similar in many aspects to the Tesla coil with the emitting loop antenna playing the role of a the Tesla primary and the receiving antenna that of the Tesla secondary coil.\\
 In fact here we are observing a splitting of the natural resonance frequency $\nu_{0}$ of the antennas into an up-shifted and a down-shifted frequency, a very general phenomena of coupled systems ranging from the hydrogen molecule to coupled optical waveguide modes.
\begin{figure}
\includegraphics[angle=0,width=7.5cm]{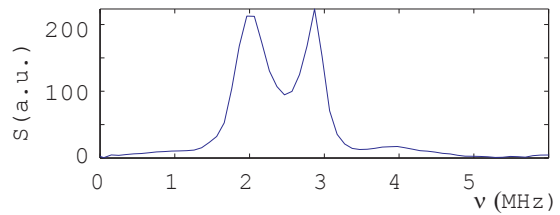}
\caption{\label{fig:fig3} Fourier transform of the V(t) trace shown in Fig.\ref{fig:fig2}.}
\end{figure}
The split frequencies are related to the natural resonance frequency through an adimensional coupling coefficient, q \cite{landau}:
\begin{eqnarray}
\nu_{1}=\frac{\nu_{0}}{\sqrt{1+q}}\\
\nu_{2}=\frac{\nu_{0}}{\sqrt{1-q}}
\end{eqnarray}
Using $\nu_{2}=1.95$ MHz and $\nu_{2}=2.88$ MHz as found from Fig.\ref{fig:fig3} we obtain $\nu_{0}=2.28$ MHz and q = 0.37.
 It is possible to double-check the $\nu_{0}$ value by simpling exciting the antenna and capacitor (LC circuit) with a cosine signal at varying frequency and searching for the resonance condition. \\
To fit the experimental V(t) data we may then used the following phenomenological relation
\begin{eqnarray}
V(t) & = & V_{0}\cdot e^{-\alpha t}\cdot [(\cos(2\pi\nu_{2}t)+\cos(2\pi\nu_{1}t)) \nonumber \\ 
& & \cdot(1-e^{-k/t^{2}})+\cos(2\pi\nu_{2}t)\cdot e^{-k/t^{2}}]  
\end{eqnarray}
The first term describes the beating the two oscillations at $\nu_{1}$ and $\nu_{2}$. This beating term is weighed by an exponential function that depends on t and is adjusted (with the parameter k) so that at t $\sim4$ $\mu$s the second term describing a simple oscillation at $\nu_{2}$ takes over. The best fit is shown as a solid line in Fig.\ref{fig:fig2} and gives us $R=6.1$ $\Omega$ and $L=4.9$ $\mu$H. The value for $L$ is in agreement with results from a sweep-test performed with a signal generator whereas $R$ also accounts for the energy loss mediated by the antenna A.\\
\begin{figure}
\includegraphics[angle=0,width=7.5cm]{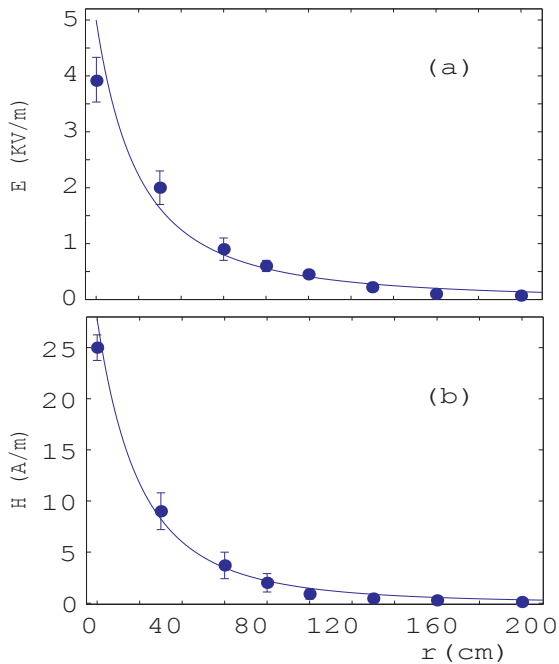}
\caption{\label{fig:fig4} (a) Electric field intensity versus distance measured along the loop antenna axis: the solid line shows the $1/r^{2}$ fit. (b) Magnetic field intensity versus distance: the solid line shows the $1/r^{3}$ fit}
\end{figure}
Finally we note that it is not possible to easily reproduce all of Hertz' results with this setup, in particular it is not feasible to try measuring the wavelength. However, on the contrary this apparatus is ideal for characterizing the near field decay rate (e.g. along the axis of the antenna A) of the EM field emitted from a loop antenna. It is expected that the near field ($r\leq \lambda$) should decay much faster than in the far field. In particular the magnetic field should decay as $1/r^{3}$ and the electric field as $1/r^{2}$. By removing RA and measuring the E and B fields along the antenna axis using an EM field meter (Wandell \& Goltermann) we obtain the results shown in Fig.\ref{fig:fig4} ((a) for the E-field and (b) for the B field). The dots show the experimental data and the solid curves are the best fit to the predicted behaviour. As can be seen the data reproduce the $1/r^{2}$ and $1/r^{3}$ dependencies quite well. However we note that it has been predicted that the E-field goes to zero for r = 0 whereas we find that although there is a certain deviation from the $1/r^{2}$ curve, E is still far from null. This may be the result of an averaging effect due to the relatively large size of the meter sensor or maybe (and most likely) due to spurious reflections and contributions coming from the surrounding environment.\\
In conclusion we have demonstrated a modified version of the original experiments performed by H. Hertz in 1888. The apparatus is relatively simple to construct but nevertheless is able to give an insight into many physical mechanisms such as RLC circuit oscillation and frequency splitting due the coupling between two resonant modes. The same experiment used in a Physics laboratory class has proved to be an interesting way to introduce students to the use of the Fourier transform and to some physical concepts (such as energy splitting) that are often treated within relatively more complex (and thus more difficult to study) contexts.



\end{document}